\documentclass[aps,prd,floats,floatfix,amssymb,amsmath,amsfonts,twocolumn,superscriptaddress,nofootinbib]{revtex4-1}

\usepackage{mathrsfs}
\usepackage{graphicx}
\usepackage{hyperref}
\usepackage{booktabs}

\newcommand{\eq}[1]{Eq.~(\ref{#1})}

\begin{document}
	
	\title{Analog model for the BTZ black hole}
	
	\author{Christyan C. de Oliveira}
	\email{chris@ifi.unicamp.br}
	\affiliation{Instituto de Física ``Gleb Wataghin'', Universidade Estadual de Campinas, 13083-859, Campinas, SP, Brazil}
	\author{Ricardo A. Mosna}
	\email{mosna@unicamp.br}
	\affiliation{Departamento de Matem\'atica Aplicada, Universidade Estadual de Campinas, 13083-859, Campinas, SP, Brazil}

	\begin{abstract}
		We present an analog model for the Ba\~nados, Teitelboim, Zanelli (BTZ) black hole based on a hydrodynamical flow. We numerically solve the fully nonlinear hydrodynamic equations of motion and observe the excitation and decay of the analog BTZ quasinormal modes in the process. We consider both a small perturbation in the steady state configuration of the fluid and a large perturbation; the latter could be regarded as an example of formation of the analog (acoustic) BTZ black hole.
	\end{abstract}
	
	\maketitle
	
	\section{Introduction}
	\label{sec:intro}
	In 1981, W. G. Unruh presented a new way of interpreting general relativity (GR) in terms of systems belonging to other areas of physics~\cite{unruh1981experimental}. This finding opened the way to the discovery of a multitude of systems that exhibit effects with close GR counterparts~\cite{barcelo2011analogue,novello2002artificial}. The study of such systems is now generically known as analog gravity.
	
	The analog gravity framework dwells on the fact that disturbances on background states of certain nongravitational systems are governed by equations of motion which are identical to those of (classical/quantum) fields in curved spacetimes. This allows several aspects of GR, which are not amenable to be directly probed, to be experimentally tested. In fact, many ideas proposed in analog gravity have been tested in the last decades. For instance, rotational superradiance \cite{torres2017rotational}, cosmological expansion \cite{eckel2018rapidly}, the ringdown of a black hole \cite{torres2019analogue,torres2020quasinormal}, and Hawking radiation have been observed \cite{rousseaux2008observation,weinfurtner2011measurement,euve2016observation,steinhauer2016observation,de2019observation,kolobov2021observation}. It is worth mentioning that the analog gravity framework does not allow one to probe dynamical aspects of the theory (those related to the Einstein equations), since the equations of motion governing the analog model have a fundamentally distinct nature. In other words, analog gravity models are only concerned with kinematical aspects of curved spacetimes.

	A particular hydrodynamical analog model for a class of spherically symmetric metrics (which include the Schwarzschild and Reissner-N\"{o}rdstrom spacetimes) has been recently proposed in Ref.~\cite{de2021analogue}. That model requires a careful fine-tuning of physical parameters of the flow, namely its local velocity and its sound speed. The local fluid velocity can be directly set up, in principle, by applying a suitable external force to the fluid. On the other hand, the local speed of sound is much less amenable of external control since it is determined by the relevant equation of state, which describes the internal forces in the flow and depends on the nature of the fluid. 
	
	In this paper we apply the ideas of~\cite{de2021analogue} but now we do not start by fixing the spacetime we want to emulate. Instead, we start by assuming that the equation of state for the fluid is as simple as it gets, so that the local speed of sound is constant throughout the fluid, see \eq{eq:eqofstate}. Interestingly enough, the curved spacetime that results from this procedure is the celebrated BTZ spacetime introduced by Ba\~nados, Teitelboim and Zanelli in~\cite{banados1992black}. 
	The BTZ spacetime is a black hole solution of $(2+1)$-dimensional GR with negative cosmological constant, which is asymptotically anti-de Sitter and has no curvature singularity at the origin (for a review, see \cite{carlip19952+}). Because of its geometrical simplicity, it has been widely used as a lower dimensional model to investigate several effects related to the foundations of classical and quantum gravity, such as  microscopic properties of black holes \cite{carlip1995lectures}. 
	
	An important characterizing property of black holes is how they respond to perturbations in the metric.  Upon perturbation, a black hole goes, in general, through a transient stage that depends on the source of the perturbation. After that, the system can be characterized by a spectrum of complex frequencies called quasinormal frequencies that depend only on the black hole parameters \cite{kokkotas1999quasi,berti2009quasinormal,konoplya2011quasinormal}. The corresponding {\it quasinormal modes} (QNMs) describe the characteristic ringdown that occurs as a response to the perturbation. The QNMs are usually defined as the modes satisfying ingoing boundary conditions at the black hole horizon and outgoing boundary conditions at infinity. This definition works perfectly fine for asymptotically flat spacetimes (Schwarzschild and Kerr black holes, for instance). However, the situation is subtler in the case of asymptotically curved spacetimes. In part, this results from the difficulty in distinguishing ingoing and outgoing waves at infinity. Moreover, for an asymptotically anti-de Sitter spacetime, the lack of global hyperbolicity gives rise to another issue: the initial conditions are not sufficient to uniquely determine the time evolution of a field, and extra boundary conditions at spatial infinity are required \cite{wald1980dynamics,ishibashi2003dynamics,ishibashi2004dynamics}. These boundary conditions influence all types of wave phenomena \cite{de2022boundary,dappiaggi2018superradiance,ferreira2017stationary}, including, in particular, the quasinormal modes.
	
	In this work we are interested in analyzing the characteristic quasinormal decay of the BTZ black hole in terms of the analog nonlinear phenomenon that takes place in the fluid as a response to perturbing its flow. Our goal is therefore to use an ideal fluid to probe the quasinormal decay of a scalar field in BTZ via the observation of the decay rate of sound waves.
	
	This paper is organized as follows. In Sec. \ref{sec:analogue BTZ black hole} we find the flow background parameters corresponding to the emulation of the BTZ spacetime by an effective metric. In Sec. \ref{sec:time evolution} we numerically solve the equations of fluid dynamics for a small perturbation in the velocity field propagating on the background found in Sec. \ref{sec:analogue BTZ black hole}.
	Using the known BTZ quasinormal frequencies \cite{cardoso2001scalar}, we show that the field intermediate-time and the late-time behaviors are well described by a superposition of QNMs. After that, we consider an example of formation of an analog BTZ black hole and use this fully nonlinear process to observe the excitation and decay of the analog BTZ quasinormal modes. Finally, Sec. \ref{sec:Discussion} is dedicated to a discussion and a brief summary of our results.

	\section{analog BTZ black hole}
	\label{sec:analogue BTZ black hole}
	We consider an inviscid barotropic fluid flowing in two spatial dimensions. Let $x$, $y$, and $t$ be the spatial and time coordinates with respect to an inertial frame of reference in the laboratory. Following~\cite{de2021analogue}, we start with a stationary one-dimensional velocity profile given by
	\begin{align}
	\vec{v}(x,y) = v(x) \hat{x}.	
	\label{eq:velocity field}
	\end{align}
	The continuity equation then implies that 
	\begin{align}
	\rho(x) = \frac{k}{|v(x)|},
	\label{eq:density}
	\end{align}
	where $k$ is a constant.
	
	The analog gravity framework is based on the fact that the wave equation for a massless scalar field, 
	\begin{align}
	\Box \phi = 0,
	\label{eq:equation of motion}
	\end{align}
	is identical to the equation of motion for sound waves in the background of a flowing fluid, with the perturbation in the velocity being given by $\delta v = -\nabla \phi$. The d'Alembert operator $\Box$ is calculated with respect to the effective metric \cite{barcelo2011analogue,visser1998acoustic} 
	\begin{equation}
	ds^2=\frac{\alpha^{2}k^ 2}{c^ 2v^ 2}\left[-(c^2-v^2)dt^2-2vdtdx+dx^2+dy^2\right],   
	\label{eq:effective metric}
	\end{equation}
	which is determined by the background flow configuration, with $c$ as the local speed of sound. The constant $\alpha$ was introduced for convenience in order to make the factor $(\alpha^{2} k^{2}/c^{2} v^{2})$ dimensionless.
	
	Let us define a new timelike coordinate 
	\begin{equation}
	\label{eq_for_T}
	T= t+ \int{\frac{v(x')}{c^2(x')-v^2(x')}dx'},
	\end{equation}
	so that the metric becomes diagonal
	\begin{align}
	ds^{2} = \frac{\alpha^{2}k^ 2}{c^ 2(x) v^ 2(x)} \left\{-\left[c^{2}(x)-v^2(x) \right]dT^2+ \right. \nonumber\\
	\left. \frac{c^{2}(x)}{c^{2}(x)-v^{2}(x)}dx^2 +dy^2\right\}. 
	\label{eq:effective metric 0}
	\end{align}
	Following~\cite{de2021analogue} we define an angular coordinate $\Theta = y/L  \mbox{ (mod $2\pi$)}$, where $L$ is a characteristic length of the analog model, and a radial coordinate\footnote{Since the velocity $v(x)$ can be positive or negative, we choose the sign in \eqref{eq:radial intermediate} so that $R(x)$ is always positive.}
	\begin{equation}
	R(x) = \pm \frac{\alpha k L}{c(x)\, v(x)},
	\label{eq:radial intermediate}
	\end{equation}
	which was chosen as the function that multiplies the resulting $d\Theta^2$ in \eq{eq:effective metric 0}. 
	In terms of the new coordinates $(T,R,\Theta)$ the metric now reads
	\begin{align}
	ds^{2} & = - \left[-\frac{\alpha^{2} k^{2} }{c^{4}(x)}  + \frac{R^{2}(x)}{L^{2}}\right]dT^{2}  \nonumber \\
	&  + \frac{R^{2}(x)/L^{2}}{\left[1 - \frac{\alpha^{2} k^{2} L^{2}}{c^{4}(x) R^{2}(x)} \right] R^{'2}(x)} dR^{2} + R^{2}(x) d\Theta^{2},
	\end{align}
	where $R'(x) = dR/dx$. We now demand that this metric be in the Schwarzschild gauge, so that $g_{11} =-\kappa^2/g_{00}$. This requires that $R(x)$ obey the differential equation 
	\begin{equation}
	\label{eq:Rlinha}
	R'^{2}(x) = \frac{c^2(x) R^{4}(x) }{\kappa^2 L^{4}}.
	\end{equation}
	
	Up to here, the argument is valid for a generic (position-dependent) speed of sound. However, differently from \cite{de2021analogue}, where we considered position-dependent speed of sound configurations (with their ensuing contrived equations of state), here we will analyze the simpler case of a constant speed of sound. In this case \eq{eq:Rlinha} can be immediately integrated to yield (up to a trivial translation in $x$)
	
	\begin{align}
	R(x) = - \frac{L^{2}}{x},
	\label{eq:radial btz}
	\end{align}
	where we took, for simplicity, $\kappa=c$ and we chose the negative sign at the right-hand side so that $R(x)$ is positive and increasing for $x\in(-\infty,0)$.
	
	As a result, the effective metric takes the form
	\begin{align}
	ds^2=-\left(- \frac{\alpha^{2} k^{2}}{c^{4}} + \frac{R^{2}}{L^{2}}  \right)dT^{2} + \left(- \frac{\alpha^{2} k^{2}}{c^{4}} + \frac{R^{2}}{L^{2}}  \right)^{-1} dR^{2}\nonumber \\
	+ R^{2} d\Theta^{2}.
	\label{eq:metric btz analogue}
	\end{align}
	We recognize (\ref{eq:metric btz analogue}) as the metric of a static BTZ black hole  with mass $M=\alpha^{2} k^{2}/c^{4}$ and curvature radius $l = L$ \cite{banados1992black}. We note that the horizon ($R=R_h:=l \sqrt{M}$) and conformal boundary ($R=\infty$) of the BTZ spacetime are realized at $x=-L/\sqrt{M}$ and $x=0$, respectively, in this model. We will also denote the boundary $x=0$ of the laboratory by $\cal{E}$. Notice that the constant $M$ is dimensionless in this spacetime.

	\section{Time evolution and analog quasinormal decay}
	\label{sec:time evolution}

	The equations of motion for an inviscid fluid subjected to an externally imposed body force $\vec{f}$ are given by~\cite{barcelo2011analogue,visser1998acoustic}
	\begin{align}
	\frac{\partial \rho}{\partial t} + \nabla \cdot \left( \rho \vec{v} \right) &=0, \quad \mbox{(continuity equation)} \label{eq:continuity equation general} \\
	\rho \left[ \frac{\partial \vec{v}}{\partial t}  + \left( \vec{v} \cdot \nabla \right) \vec{v}\right] &= -\nabla p +  \vec{f}, \quad \mbox{(Euler equation)} \label{eq:euler equation general} 
	\end{align}
	where $p$ is the pressure, which here satisfies the equation of state
	\begin{equation}
	\label{eq:eqofstate}
	p = c^{2} \rho,
	\end{equation}
	with constant $c$, as discussed above.
	
	We are concerned with two-dimensional flows with physical quantities varying along $x$ only.
	More explicitly, density and pressure will depend only on $x$ [i.e., $\rho = \rho(x)$, $p = p(x)$], the velocity $\vec{v}$ will be given by \eqref{eq:velocity field} and the external force density will be given in terms of a driving potential $\Phi(x)$,  
	\begin{align}
	\vec{f}(x) = -\rho  \nabla \Phi =- \rho  \partial_{x} \Phi \, \hat{x}.
	\end{align}
	The external potential is taken to be fixed, which means that it is insensitive to backreaction, as in  \cite{visser1998acoustic,barcelo2011analogue}. Therefore, the discussion of \cite{bilic2022analog} does not apply to the present work (nor to~\cite{de2021analogue}). 
	With the assumptions made above, the equations of motion simplify to
	\begin{align}
	\partial_{t} \rho + \partial_{x} \left( \rho v \right) &=0, \label{eq:continuity equation}\\
	\rho \left( \partial_{t} +v \partial_{x} \right)v &= - \partial_{x}p - \rho \partial_{x} \Phi  \label{eq:euler equation}.
	\end{align}
	
	The fluid configuration that implements the results of the previous section can be obtained from Eqs.~\eqref{eq:density}, \eqref{eq:radial intermediate} and \eqref{eq:radial btz}, which determines the background velocity
	\begin{equation}
	\label{eq:v0}
	v_0(x) = \left(\frac{\alpha k}{cL}\right) x,
	\end{equation}
	and the background density
	\begin{equation}
	\label{eq:rho0}
	\rho_0(x) = - \left(\frac{c L}{\alpha}\right) \frac{1}{x}.
	\end{equation}
	From the Euler equation \eqref{eq:euler equation}, we find the external potential 
	\begin{align}
	\Phi(x) = c^{2} \log \left(\frac{x}{L}\right) - \left(\frac{\alpha^2 k^{2}}{c^2 L^2}\right) \frac{x^2}{2}.
	\label{eq:external potential}
	\end{align}
	
	We now consider perturbations of the above steady-state configuration of the fluid and follow the evolution of the relevant physical quantities in time. In order to do that, we numerically solve the nonlinear fluid equations and compare the result with what would have been the corresponding evolution on the BTZ black hole. As we show in the following, the propagation of the fluid in this regime allows one to recover the mechanism of excitation of quasinormal modes at the black hole level. We do this for both a small perturbation and a large perturbation; the latter could be regarded as an example of the process of formation of an acoustic BTZ black hole. In all the cases the acoustic black hole has its quasinormal modes excited at the well-known quasinormal frequencies of the BTZ black hole.

	\subsection{The Cauchy problem}
	\label{sec:Cauchy problem}
	
	From the analog model perspective, we want to emulate a massless scalar field propagating on a BTZ background. The corresponding Cauchy problem in this spacetime is given by the differential equation
	\begin{align}
	\Box \phi& = 0, \label{eq:cauchy diff eq} 
	\end{align}
	along with the initial data 
	\begin{align}
	\phi|_{t_{0}} = \phi_{0},  & & \partial_{t} \phi |_{t_{0}} = \dot{\phi}_{0}, 
	\label{eq:cauchy initial conditions} 
	\end{align}
	where $\phi_{0}$, $\dot{\phi}_0$ are smooth functions defined on the surface $t = t_0$.  
	
	This problem has some particular features that are worth mentioning. First, since the BTZ black hole is a nonglobally hyperbolic spacetime, in general the time evolution determined by the differential equation \eqref{eq:cauchy diff eq} together with initial data \eqref{eq:cauchy initial conditions} is not well defined.%
	\footnote{See \cite{wald1980dynamics,ishibashi2003dynamics,ishibashi2004dynamics} for the general theory of field dynamics in nonglobally hyperbolic spacetimes. For the case of the BTZ black hole, see \cite{garbarz2017scalar}.}
	Physically, the lack of global hyperbolicity is related to the fact that information traveling in spacetime can reach (or come from) spatial infinity in a finite time. Therefore, some extra boundary condition at spatial infinity is required to ensure a unique physically sensible time evolution for the field $\phi$.
	
	It follows from the field theory on BTZ black holes \cite{de2022boundary,ferreira2017stationary,dappiaggi2016hadamard} that the radial part of the field 
	\begin{align}
	\phi(T,R,\Theta) = \frac{\psi(R)}{\sqrt{R}} e^{-i \omega T} e^{i m \Theta}
	\end{align} 
	can be expressed as a linear combination 
	\begin{align}
	\psi = C_{D} \psi^{(D)} + C_{N} \psi^{(N)},
	\end{align}
	of two linearly independent solutions $\psi^{(D)}$, $\psi^{(N)}$. The function $\psi^{(D)}$ is chosen to be the principal solution at $R\to \infty$, that is, the unique solution (up to scalar multiples) such that 
	\begin{align}
	\lim_{R\to \infty}\frac{\psi^{(D)}(R)}{g(R)} =0,
	\end{align}
	for every solution $g(R)$ not proportional to $\psi^{(D)}$. The function $\psi^{(D)}$ is also called the generalized Dirichlet solution.
	The other solution, $\psi^{(N)}$, is a nonprincipal solution and it is not unique, since any linear combination of this function with the principal solution is another nonprincipal solution. $\psi^{(N)}$ is also called a generalized Neumann solution.
	For the massless scalar field, only the solution $\psi^{(D)}$ is square integrable at $R\to \infty$ \cite{ferreira2017stationary}. Hence, the field has to satisfy a Dirichlet boundary condition at spatial infinity
	\begin{equation}
	\phi|_{R=\infty} =0.
	\label{eq:dirichlet bc}
	\end{equation} 
	
	In particular, a quasinormal mode will be characterized by ingoing boundary condition at the black hole horizon and Dirichlet boundary condition at infinity. These modes are given by \cite{cardoso2001scalar}
	\begin{align}
	\psi(R(z)) & =(1-z)^{-1/4}z^{-\frac{i \omega }{2}} \times &  \nonumber \\
	& F\left(-\frac{i m}{2}-\frac{i \omega }{2},\frac{i m}{2}-\frac{i \omega }{2};1-i \omega ;z\right),
	\end{align}
	where $z = 1 - R^{2}_{h}/R^{2}$,  $F(\alpha,\beta,\gamma,z)$ stands for the standard hypergeometric function and the quasinormal frequencies are given by
	\begin{equation}
	\omega_{nm} = \pm m -2i(n+1), \qquad n, \, m  = 0, 1, 2,3, \dots,
	\label{eq:qnm frequencies}
	\end{equation}
	where $m$ is the angular quantum number and $n$ labels the imaginary part of the frequency, which is related to the characteristic time of decay of the corresponding mode.
	
	Notice that these modes satisfy  
	\begin{align}
	\phi|_{R=\infty} =0, & & \frac{\partial \phi}{\partial R} \bigg|_{R=\infty} = 0.
	\end{align}
	On the analog model end, these boundary conditions correspond to vanishing perturbation in the velocity and density profiles at $x=0$.
	
	A nice feature of our model is that it maps the black hole spatial infinity to the physical (finite) boundary $\cal{E}$ of the system at the laboratory, at $x=0$. Hence, the boundary conditions at $x=0$ that are required by the sound propagation in the fluid can be naturally chosen to emulate the massless scalar field in the BTZ spacetime. For the fluid motion, these boundary conditions ensure that the energy flux across the boundary $\cal{E}$ is zero. At the spacetime level, these boundary conditions mean that information can neither escape to nor come from the spatial infinity.
	
	\subsection{Small perturbation and QNM excitation}
	\label{sec:small perturbation}
	Let us consider as initial conditions a configuration for which $v$ is slightly perturbed from the steady state configuration $v_0$ at a given point $x_0$:
	\begin{align}
	v(t=0,x) = v_{0}(x) + \delta v(x), \\
	\rho(t=0,x) = \rho_{0}(x) + \delta \rho(x),
	\end{align}
	with 
	\begin{align}
	\delta v(x) & = A \, e^{- \frac{\left(x-x_{0}\right)^{6}}{2 \sigma^{2}}}, \label{eq:initial data 1}\\
	\delta \rho(x) & = 0. \label{eq:initial data 2}
	\end{align}
	
	We choose units such that $\alpha=k=c=1$; the black hole mass then becomes $M= 1$. For simplicity, we also choose the width $L$ as $L=1$. The exterior region of the black hole is then mapped into the interval $-1 < x < 0$, with $x=-1$ corresponding to the horizon, and $x=0$ corresponding to spatial infinity.

	To simulate Dirichlet boundary conditions at infinity, we should impose that the disturbance vanishes at $x=0$,
	\begin{align}
	\delta v(x=0) =0, \\
	\delta \rho(x=0) =0.
	\end{align}
	In order to avoid numerical difficulties, we impose boundary conditions at $x=-\epsilon$, with $\epsilon>0$ being a sufficiently small parameter, instead of at $x=0$. More explicitly, we require 
	\begin{align}
	v(t,x=-\epsilon) & = v_0(-\epsilon), \label{veps}\\ 
	\rho(t,x=-\epsilon) & = \rho_{0}(-\epsilon).  \label{rhoeps}
	\end{align}
	
\begin{figure}[th!]
		\centering
		\includegraphics[width=\columnwidth]{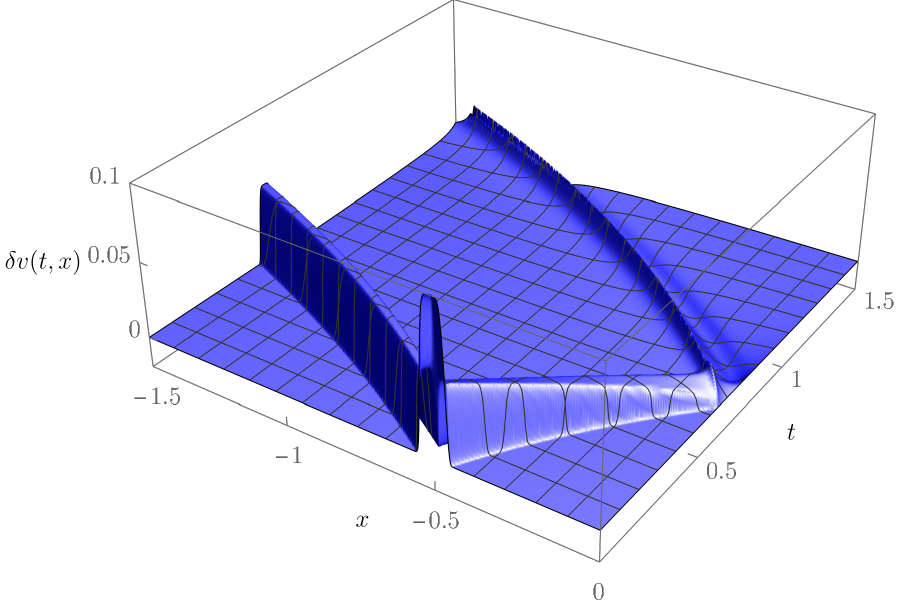}
		\includegraphics[width=\columnwidth]{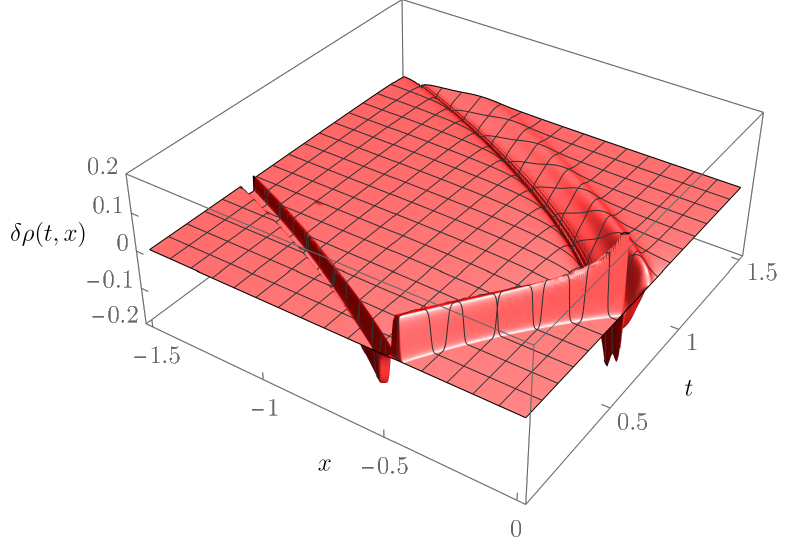}
		\caption{	
			Time evolution of a initial perturbation in the background velocity (top) and density (bottom) given by Eqs.~(\ref{eq:initial data 1}) and (\ref{eq:initial data 2}). The parameters were chosen as $\epsilon = 10^{-7}$, $A = 0.1$, $\sigma = 0.00005$, and $x_{0} = -0.5$. The perturbation splits into two portions. One goes towards $x=0$ and is reflected at time $t \sim 0.6$. The other goes towards the horizon and falls into the supersonic region $(x<-1)$.   
		}
		\label{fig:v de t x}
	\end{figure}

	We solved the system given by Eqs.~\eqref{eq:continuity equation} and \eqref{eq:euler equation} for $v(t,x)$ and $\rho(t,x)$ with the software \textit{Mathematica}~\cite{Mathematica}.%
	\footnote{We used its NDSolve routine with a MaxStepSize set to $0.001$. Our calculations showed good numerical convergence, with the results being the same for values of $\epsilon$ ranging from $10^{-3}$ to $10^{-7}$.} 
	Figure~\ref{fig:v de t x} shows the obtained time evolution of a perturbation initially centered at $x_{0}=-0.5$. 	We see that the initial disturbance splits into two portions: one goes towards the horizon and falls into the supersonic region $(x<-1)$. The other goes towards $x=0$ and, around $t \sim 0.6$, is reflected at the boundary and redirected towards the horizon. Although the expression of the analog metric is degenerate at the horizon (as it occurs for a Schwarzchild black hole, for instance), the fluid physical quantities and their corresponding perturbations are both well defined there. We see that these physical quantities are also well defined in the supersonic region ($x<-1$).
	
	\subsubsection{QNM decay}
	
	It follows from the general theory of wave propagation on black hole spacetimes that the response to a perturbation on the background geometry has, in general, three distinct stages \cite{leaver1986spectral,ching1995wave}: (i) the early time response, which depends highly on the initial conditions; (ii) the intermediate-time regime, which is dominated by a QNM ringing; and (iii) the late-time regime, which is governed by a power law tail. Mathematically, the quasinormal modes arise from the poles of the Green's function associated with the wave equation, and the power law tail comes from a branch cut on the Green's function domain.

	However, differently from the Schwarzschild and Kerr black holes, where a branch cut on the Green's function frequency domain gives rise to a late-time power law tail \cite{leaver1986spectral,andersson1995excitation,andersson1997evolving}, the Green's function associated with wave propagation on the BTZ black hole has no branch cut on the $\omega$-complex plane. This results in a late-time (exponential) decay governed by the quasinormal ringing \cite{chan1997scalar}.
	
	In the following, we fit the intermediate and late-time behavior of our numerical solution, at a fixed position of observation $x_{obs}$, to a linear superposition of the first $N$ quasinormal modes~\cite{okuzumi2007quasinormal}. We take
	\begin{align}
	u (\mathbf{C}; t) = v_{0}(x_{obs}) +  \sum_{n=0}^{N} C_{n} e^{-i \omega_{n0} (t-t_{1})},
	\label{eq:fitting function}
	\end{align} 
	where $\omega_{n0}$ are the frequencies given by \eqref{eq:qnm frequencies} and $\mathbf{C} = (C_{0}, C_{1},C_{2}, \dots , C_{N})$ are fitting parameters. 
	We note that only frequencies with $m=0$ are excited since nothing depends on the analog angular coordinate $y = \Theta L$.%
	\footnote{We note that modes with nontrivial angular dependence ($m\ne0$) cannot be considered in this model unless we impose periodic boundary conditions identifying the lines $y=0$ and $y=L$.}
	We find the quasinormal approximation by minimizing the integral
	\begin{align}
	E(\mathbf{C}) = \int_{t_{1}}^{t_{2}}\left[ v(t,x_{obs}) - u (\mathbf{C}; t) \right]^{2}dt.
	\label{eq:erro}
	\end{align}
	The time interval $(t_1,t_2)$ should be chosen within the time domain where the numerical solution $v(t,x)$ is dominated by the QNM decay.

\begin{center}
		\begin{figure}[th!]
			\centering
			\includegraphics[width=\columnwidth]{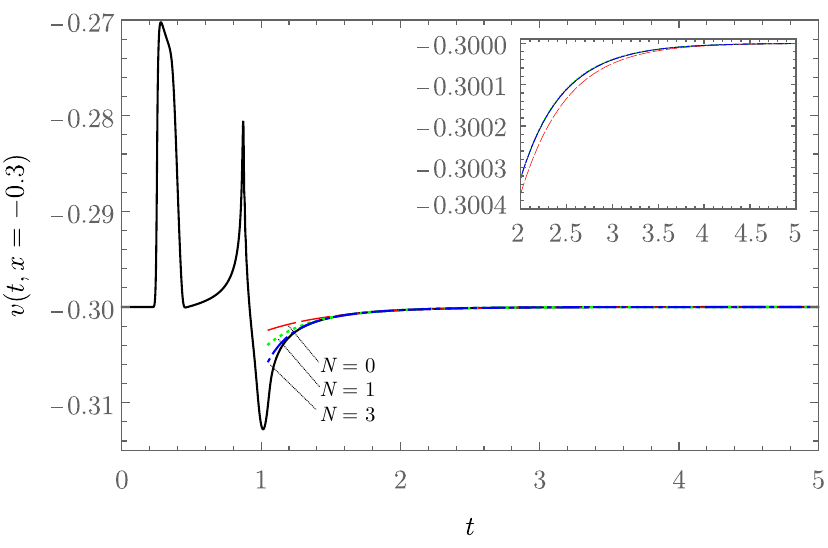}
			\caption{Numerical waveform $v(t,x=-0.3)$ (black solid curve) and quasinormal modes for a perturbation on the analog BTZ background. The parameter $\epsilon$ was chosen as $10^{-7}$.
				Top right: quasinormal approximation to late-time behavior. The red dashed curve represents the least-damped mode ($N=0$), the green dotted curve represents the sum of the first two modes ($N=1$), and the blue dot-dashed curve represents the sum of the first four modes ($N=3$). The integral \eqref{eq:erro} was calculated with $t_1 = 1.5$ and $t_2 = 5$.}
			\label{fig:v de t}
		\end{figure}
	\end{center}

	Figure~\ref{fig:v de t} shows the numerical velocity profile (solid curve) at fixed position $x=x_{obs}=-0.3$ as a function of time. We see the initial perturbation passing through the observation point around $t \sim 0.3$. The reflected pulse comes around $t\sim0.85$. After $t \sim 1$ the  quasinormal modes govern the signal decay.
	We also see in Fig.~\ref{fig:v de t} the quasinormal fitting obtained from Eq.\eqref{eq:fitting function} for $N=0$ (red dashed curve), $N = 1$ (green dotted curve), and $N=3$ (blue dot-dashed curve). The corresponding parameters $C_{n}$ are listed in Table \ref{tab:fitting parameters 1}.
	
	\begin{table} 
		\centering
		\caption{Parameters $C_n$ for the quasinormal approximation on the scenario of a small perturbation on a steady background flow.}
		\begin{tabular}{l c c lclcl}
			\toprule[.5pt]
			\midrule[.5pt]
			& & &  $N=0$             &  &    $N= 1$       &  & $N = 3$  \\
			\midrule[.5pt] 
			\midrule[.5pt]
			$C_0$ & &  &     -0.000983912  &  & -0.000757182    &  & -0.00078922 \\
			$C_1$ & &  &                   &  & -0.000340095    &  & -0.000230737    \\
			$C_2$ & &  &                   &  &                 &  & -0.0000441535  \\
			$C_3$ & &  &                   &  &                 &  & -0.0000469776     \\
			\bottomrule[.5pt]
		\end{tabular}
		\label{tab:fitting parameters 1}
	\end{table}

	\subsection{Large perturbation: Acoustic black hole formation}
	\label{sec:Transition to an acoustic black hole state}
	
	As another example of excitation of quasinormal modes, we now consider a possible scenario for the formation  of the acoustic BTZ black hole. We set, as initial state for the system, a particular configuration where the fluid starts with zero velocity everywhere and let it evolve while subjected to the same external potential given by Eq.~\eqref{eq:external potential}. Here it is worth recalling that the analog gravity framework can only probe kinematical aspects of GR, as opposed to dynamical aspects emerging from the Einstein field equations. As such, the model presented in this section does not emulate the actual dynamical evolution of the BTZ spacetime metric. The purpose of this example is to illustrate one possible formation process for the analog BTZ black hole and to analyze the corresponding excitation of its quasinormal modes.

\begin{figure}[th!]
		\centering
		\includegraphics[width=\columnwidth]{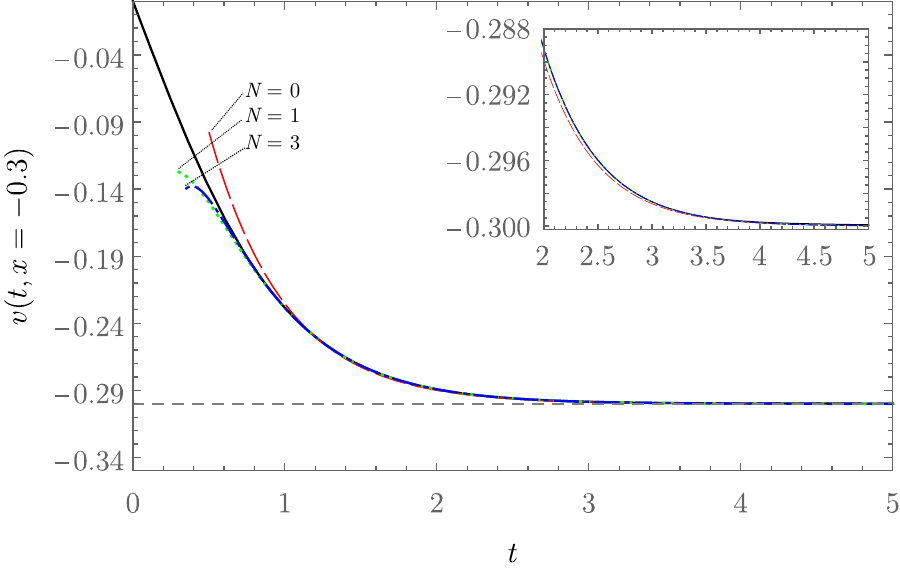}
		\caption{Numerical waveform $v(t,x=-0.3)$ (black solid curve) and quasinormal modes for an example of formation of the acoustic BTZ black hole.  The parameter $\epsilon$ was again chosen as $10^{-7}$. The first phase of the transition occurs roughly between $t\sim 0.8$ and $t\sim 1.6$. After that, the  quasinormal modes govern the signal. Top right: quasinormal approximation to late-time behavior. The red dashed curve represents the least-damped mode ($N=0$), the green dotted curve represents the sum of the first two modes ($N=1$), and the blue dot-dashed curve represents the sum of the first four modes ($N=3$). The integral \eqref{eq:erro} was calculated with $t_1 = 1$ and $t_2 = 5$.}
		\label{fig:v de t form}
	\end{figure}

	To simulate this scenario numerically, we have taken the initial conditions 
	\begin{align}
	v(t=0,  x) & = 0, \\
	\rho(t=0,  x) & = \rho_{0}(x),
	\end{align}
	and solved the fluid equations (\ref{eq:continuity equation}) and (\ref{eq:euler equation}) with the software \textit{Mathematica}~\cite{Mathematica}.%
	\footnote{The boundary conditions were again given by Eqs.~(\ref{veps}) and (\ref{rhoeps}), with the results being the same for values of $\epsilon$ ranging from $10^{-3}$ to $10^{-7}$. This time we used the routine NDSolve with a MaxStepSize set to $0.005$.} 
	Figure ~\ref{fig:v de t form} shows the velocity profile at the observation point $x=-0.3$. We again found the contribution of the quasinormal modes to the waveform by using the fitting function \eqref{eq:fitting function}. The values found for $C_{n}$ are listed in Table \ref{tab:fitting parameters 2}.
	
	We observe from Fig.~\ref{fig:v de t form} that the initial phase of the transition takes place roughly between $t\sim 0.8$ and $t\sim 1.6$. After that, the  quasinormal modes govern the signal. We also see the late-time behavior of the velocity field (black solid curve) together with quasinormal profiles for the least-damped mode, $N=0$ (red dashed curve), a superposition of the first two modes, $N=1$ (green dotted curve), and of the first four modes, $N=3$ (blue dot-dashed curve). After the quasinormal regime, $t\gtrsim4$, the flow approximately  reaches equilibrium at the steady state configuration of the acoustic BTZ black hole.  
	
	\begin{table}[h] 
		\centering
		\caption{Parameters $C_n$ for the quasinormal approximation on the scenario of formation of the analog BTZ black hole.}
		\begin{tabular}{l c c lclcl}
			\toprule[.5pt]
			\midrule[.5pt]
			& & &  $N=0$                   &  &    $N= 1$     &  & $N = 3$  \\
			\midrule[.5pt] 
			\midrule[.5pt] 
			$C_0$ & &  &     0.074547     &  &  $\,\, 0.0808317$    &  &  $\,\, 0.0812623 $     \\
			$C_1$ & &  &                   &  & -0.00942706   &  & -0.0113435       \\
			$C_2$ & &  &                   &  &               &  &  $\,\, 0.00208244 $    \\
			$C_3$ & &  &                   &  &               &  &  -0.000485368   \\
			\bottomrule[.5pt]
		\end{tabular}
		\label{tab:fitting parameters 2}
	\end{table}
	
	\section{conclusion}\label{sec:Discussion}	
	
	We proposed an analog model for the BTZ black hole based on a  unidirectional flow of a nonhomogeneous fluid. We have considered a barotropic fluid obeying a simple equation of state, which corresponds to a constant local speed of sound. The physical quantities describing the flow vary along just one direction. In particular, the flow velocity field points to a fixed direction in the laboratory reference frame. The coordinate describing the direction of the flowing fluid is mapped into the radial coordinate of the analog spacetime. Following the steps presented in \cite{de2021analogue} we were naturally led to find the effective acoustic metric as that of the well-known BTZ black hole.
	
	A nice feature of our model is that the exterior region of the BTZ black hole is mapped into a finite region in the laboratory. In particular, the BTZ conformal boundary is mapped into the boundary $\cal{E}$ at the laboratory which is at a finite distance from the acoustic horizon. Since the BTZ black hole is a nonglobally hyperbolic spacetime, its conformal boundary plays a fundamental role in the dynamics of fields propagating on it. On the analog model end, the extra boundary condition (at the conformal boundary) required to uniquely determine the time evolution of the field can be naturally interpreted as a boundary condition for the sound propagation in the laboratory at $\cal{E}$. 
	
	Finally, we considered configurations with both small and large deviations from the steady state. In the latter case we numerically followed an example of formation of the acoustic black hole. In both cases we examined how the associated QNMs are excited. 
	
	Although the experimental realization of the analog model presented here (for which the sound waves propagate in the bulk) is beyond the scope of the present work, we would like to point out that the method of obtaining analog models used in this paper (and introduced in  \cite{de2021analogue}) can also be applied to gravity waves propagating on the surface of a liquid within a shallow basin with curved bottom. In this context, the dynamical degrees of freedom, instead of $\rho$ and $p$, are the variable fluid depth $h(x)$ and a specific nonflat bottom for the basin. This could provide an alternative framework, closer to experimental realization, wherein our method can still be applied.

	\acknowledgments
	The authors acknowledge insightful discussions with M. R. Correa, J. P. M. Pitelli and M. Richartz. C. C. O. acknowledges support from the Conselho Nacional de Desenvolvimento Cient\'{i}fico e Tecnol\'{o}gico (CNPq, Brazil), Grant No. 142529/2018-4. R. A. M. was partially supported by Conselho Nacional de Desenvolvimento Científico e Tecnológico under Grant No. 310403/2019-7.

\end{document}